\documentclass[12pt]{article}

\usepackage{cite}
\usepackage{epsf}
\usepackage{graphicx}
\usepackage{psfrag}
\usepackage[dvips]{epsfig}

\usepackage[cp866]{inputenc}
\usepackage[english]{babel}
\usepackage{amssymb,indentfirst}

\makeatletter
\renewcommand \thesection {\@arabic\c@section.}
\renewcommand\thesubsection   {\thesection\@arabic\c@subsection.}
\renewcommand\thesubsubsection{\thesubsection\@arabic\c@subsubsection.}
\makeatother

\def\lineup#1{\mbox{$\raise1.0ex\hbox{--} \kern-0.6em#1$}}

\def\starup#1{\mbox{$\raise1.6ex\hbox{$*$} \kern-0.5em#1$}}
\def\starupp#1{\mbox{$\raise1.8ex\hbox{$*$} \kern-1.0em#1$}}
\def\staruppp#1{\mbox{$\raise1.0ex\hbox{$*$} \kern-0.5em#1$}}
\def\sstarup#1{\mbox{\scriptsize $\raise1.8ex\hbox{$*$} \kern-.7em#1$}}

\def\ttildeup#1{\mbox{$\raise0.0ex\hbox{\Large $\; \tilde{}$} \kern-0.45em#1$}}
\def\bbar#1{\mbox{$\raise-0.4ex\hbox{\Large $\; \bar{}$} \kern-0.35em#1$}}

\begin{document}

\title{
Mass limits for scalar and gauge leptoquarks from $ K_L^0 \to e^{\mp} \mu^{\pm} , \; 
 B^0 \to e^{\mp} \tau^{\pm} $ decays
 }

\author{ A.D.~Smirnov\thanks{E-mail: asmirnov@univ.uniyar.ac.ru}\\
{\small\it Division of Theoretical Physics, Department of Physics,}\\
{\small\it Yaroslavl State University, Sovietskaya 14,}\\
{\small\it 150000 Yaroslavl, Russia.}}
\date{}
\maketitle

\begin{abstract}
\noindent
The contributions of scalar and gauge leptoquarks into widths of
$K^0_L \to e^{\mp} \mu^{\pm}$, $B^0 \to e^{\mp} \tau^{\pm}$ decays are 
calculated in the models with the vectorlike and chiral four color symmetry 
and with the Higgs mechanism of the quark-lepton mass splitting. 
From the current data on $K^0_L$ and $B^0$ decays the mass limits for scalar 
and chiral leptoquarks and the updated vector leptoquark mass limits 
are obtained. It is shown that unlike the gauge leptoquarks the scalar 
leptoquark mass limits are weak, of order or below their direct mass limits.   
The search for such scalar leptoquarks at LHC and the further search for 
leptonic decays $ B^0 \to l^+_i l^-_j $ are of interest.     

\vspace{5mm}
\noindent
\textit{Keywords:} Beyond the SM; four-color symmetry; Pati--Salam; 
leptoquarks.   

\noindent
\textit{PACS number:} 12.60.-i

\end{abstract}





The search for a new physics beyond the Standard Model (SM) is now one
of the aims of the high energy physics. Putting LHC into operation will essentially enlarge 
the possibilities for such search and there is a lot of the variants of new physics 
which can give new effects at energies of LHC 
(supersymmetry, left - right symmetry, two Higgs model, etc.). 

One of the possible
variants of such new physics can be the variant induced by the
possible four color symmetry \cite{PS} between quarks and leptons.
The four color symmetry can be unified with the SM by the gauge group 
\begin{eqnarray}
     G_{new}=G_c \times SU_L(2) \times U_R(1) 
\label{eq:Gnew}
\end{eqnarray}
where $G_c$ is the group of the four color symmetry. 
This group can be either the vectorlike group \cite{PS, AD1, AD2}   
\begin{eqnarray}
    G_c = SU_V(4) 
\label{eq:G4V}
\end{eqnarray}
or the group of the general chiral four color symmetry  
\begin{eqnarray}
G_c = SU_L(4) \times SU_R(4)  
\label{eq:G4LR}
\end{eqnarray}
or  one of the special groups  
\begin{eqnarray}
      G_c = SU_L(4) \times SU_R(3) , \,\,\,\, G_c = SU_L(3) \times SU_R(4)  
\label{eq:G4SLR}
\end{eqnarray}
of the left (or right) four color symmetry. 
According to these groups the four color symmetry predicts in the gauge sector either the vector 
leptoquarks or the left and right leptoquarks or the left (right) gauge leptoquarks respectively. 
The mass limits for the vector leptoquarks are well known and the most stringent of them are 
the indirect mass limits resulted from
$K^0_L \to e^{\mp} \mu^{\pm}$ decay and they are of order of $10^{3} \,\, TeV$ 
\cite{VW, KM1,KM2}. 
By this reason it is usually thought that the effects of the four color symmetry 
at the colliders energies are too small to be directly detectable 
in the collider experiments.   

It should be noted however that the four color symmetry allows also the existence of 
scalar leptoquarks and such particles have been phenomenologically introduced in ref.~\cite{BRW} 
and were discussed in a number of papers. 
The experimental lower mass limits for the scalar leptoquarks from their direct search 
are about $ 250 \, GeV $ or slightly less in dependence on some additional assumptions~\cite{PDG06}. 
As concerns the indirect mass limits for the scalar leptoquarks they depend on the magnitude 
of the scalar leptoquark coupling constants with fermions which under phenomenological introduction 
 are arbitrary so that only the relation of these coupling constants to the leptoquark masses can be 
restricted experimentally. 

Nevertheless there is the situation when the typical magnitudes of the scalar leptoquark 
coupling constants with fermions are known. 
Indeed, in the case of Higgs mechanism of the quark-lepton mass splitting 
 the four color symmetry of type (\ref{eq:Gnew})--(\ref{eq:G4V})(MQLS model \cite{AD1, AD2, ADPv}) 
predicts the $SU(2)_L$ scalar leptoquark doublets 
\begin{eqnarray}
S^{(\pm)}_{a \alpha} = \left ( \begin{array}{c}
S_{1 \alpha}^{(\pm)}\\
S_{2 \alpha}^{(\pm)}
\end{array} \right )   
\label{eq:SpSm}
\end{eqnarray}
 with Yukawa coupling constants which occur (due their Higgs origin) to be proportional to 
the ratios $m_{f}/ \eta $ of the fermion masses $m_f$ to the SM
VEV $\eta$. As a result these coupling constants are known (up to mixing parameters) 
and they are small for the ordinary $u-, d-, s-$ quarks 
(~$ m_u/ \eta \sim m_d/ \eta \sim 10^{-5},  m_s/ \eta \sim 10^{-3}$~) 
but they are more significant for $c-, b-$ quarks ($ m_c/ \eta \sim m_b/ \eta \sim 10^{-2} $) and, 
especially, for $t$-quark ( $ m_t/ \eta \sim 0.7$ ). 
 
The analysis of the contributions of these scalar leptoquark doublets into 
radiative corrections $S-, T-, U-$ parameters showed \cite{AD3,PovSm2} 
that these scalar leptoquarks can be relatively light, with masses below 1 TeV.   
Keeping in mind that the most stringent mass limits for the vector leptoquarks are resulted from
$K^0_L \to e^{\mp} \mu^{\pm}$ decay it is interesting to know what mass limits for 
the scalar leptoquarks can be extracted from $K^0_L \to e^{\mp} \mu^{\pm}$ decay 
and from other decays of such type. 

In this paper we calculate the contributions of the scalar and vector leptoquarks into 
$K_L^0 \to e^{\mp} \mu^{\pm}$ and $ B^0 \to e^{\mp} \tau^{\pm} $ decays in frame of MQLS-model 
based on the vectorlike four color symmetry (\ref{eq:Gnew})--(\ref{eq:G4V}) and discuss 
the mass limits resulted from the current data on these decays for the leptoquarks under 
consideration. 
We also calculate the contributions into these decays from the chiral gauge leptoquarks iduced by the chiral 
four color symmetry and discuss the corresponding chiral gauge leptoquark mass limits.   

In MQLS model the basic left ($L$) and right ($R$) quarks
${Q'}^{L,R}_{ia\alpha}$ and leptons ${l'}^{L,R}_{ia}$
form the fundamental quartets of the group (\ref{eq:G4V}) and can
be written, in general, as superpositions 
\begin{eqnarray}
{Q'}^{L,R}_{ia\alpha}=\sum_{j}(A^{L,R}_{Q_a})_{ij} \, Q^{L,R}_{ja\alpha} ,
\,\,\,\,\,\,\,
{l'}^{L,R}_{ia}=\sum_{j}(A^{L,R}_{l_a})_{ij} \, l^{L,R}_{ja} \; \label{eq:fmix}
\end{eqnarray}
of the quark and lepton mass eigenstates $Q^{L,R}_{ia\alpha}$ ,
$l^{L,R}_{ia}$, where $i,j=1, \, 2, \, 3$ are the generation indexes,
$ a = 1, 2 $ and $ \alpha = 1, 2, 3 $ are the $ SU_{L}(2) $  and
$ SU_{c}(3) $ indexes,
$Q_{i1} \equiv u_i=(u,c,t)$, $Q_{i2} \equiv d_i=(d,s,b)$ are
the up and down quarks, $l_{j1} \equiv \nu_{j}$
are the mass eigenstates of neutrinos and
$l_{j2} \equiv l_{j}=(e^{-}, \mu^{-}, \tau^{-})$
are the charged leptons.
The unitary matrices $A^{L,R}_{Q_a}$ and $A^{L,R}_{l_a}$ describe
the fermion mixing and diagonalize the mass matrices of quarks and
leptons.

The Higgs mechanism of the quark-lepton mass splitting needs, in
general, two scalar multiplets $\Phi^{(2)}$ and $\Phi^{(3)}$ (with
VEV $\eta_2$ and $\eta_3$) transforming according to the
representations   (1.2.1) and (15.2.1) of the group (\ref{eq:Gnew})--(\ref{eq:G4V}).
The multiplet (15.2.1) contains as a part the scalar leptoquark doublets (\ref{eq:SpSm}) 
the down components of which $S_{2 \alpha}^{(\pm)}$ have electric charges $\pm 2/3$ and 
contribute to the leptonic decays of $K_L^0$ and $B^0$ mesons.  

In general case
the scalar leptoquarks $S_{2 \alpha}^{(+)}$ and
$\starup{S_{2\alpha}^{(-)}}$
with electric charge 2/3 are mixed and can be
written as superpositions
\begin{eqnarray}
S_2^{(+)}&=&\sum_{m=0}^3 c_m^{(+)}S_m, \; \; \; \; \; \;
 \starup{S_2^{(-)}}=\sum_{m=0}^3 c_m^{(-)}S_m \label{eq:mixS}
\end{eqnarray}
of three physical scalar leptoquarks $S_1$, $S_2$, $S_3$ with
electric charge 2/3 and a small admixture of the Goldstone mode
$S_0$. Here $c^{(\pm)}_{m}$, $m=0,1,2,3$ are the elements of the
unitary scalar leptoquark mixing matrix, $|c^{(\pm)}_{0}|^2=\frac
{1} {3}g_4^2 \eta_{3}^{2}/m_V^2 \ll 1$, $g_4$ is the $SU_V(4)$
gauge coupling constant, $\eta_3$ is the VEV of the
(15,2,1)-multiplet and $m_V$ is the vector leptoquark mass.

In particular case of the two leptoquark mixing the superpositions~(\ref{eq:mixS}) 
can be approximately written as  
\begin{eqnarray}
S^{(+)}_2 = c S_1 + s S_2, 
\nonumber \\
\starup{S^{(-)}_2} = -s S_1 + c S_2  
\label{eq:mixS2}
\end{eqnarray}
where $c = cos\theta, s = sin\theta, $ $ \theta $ is the scalar leptoquark mixing angle. 

The interaction of the vector and scalar leptoquarks with 
down fermions can be described by the lagrangians 
\begin{eqnarray}
  \emph{L}_{Vdl} &=& \frac{g_4}{\sqrt{2}} (\bar{d}_{p \alpha} [(K^{L}_2)_{pi} \gamma^{\mu}P_L +  
(K^{R}_2)_{pi} \gamma^{\mu}P_R ] l_i) V_{\alpha \mu} + h.c. ,
\label{eq:lagrVdl}\\
  \emph{L}_{Sdl} &=& (\bar{d}_{p \alpha} [ (h^L_m)_{pi} P_L +  
(h^R_m)_{pi} P_R ] l_i) S_{m {\alpha}} + h.c. 
\label{eq:lagrSdl}
\end{eqnarray}
where  $g_4 = g_{st}(M_c)$ is the $SU_V(4)$ gauge coupling constant related 
to the strong coupling constant at the mass scale $M_c$ of the $SU_V(4)$ symmetry breaking 
and $(h^{L,R}_m)_{pi}$ are Yukawa coupling constants, 
$p, i = 1,2,3, $... are the quark and lepton generation idexes, 
the index $ m $ numerates the scalar leptoquarks, 
$\alpha=1,2,3$ is the $SU(3)$ color index and  
$P_{L,R}=(1\pm \gamma_5)/2$ are the left and right operators of fermions. 
The unitary matricies $K^{L,R}_a = (A^{L,R}_{Q_a})^+ A^{L,R}_{l_a}, a=1,2$  describe 
the (down for $a=2$) fermion mixing 
in the leptoquark currents and in general case they can be nondiagonal. 
These four matrices are specific for the model with the four color quark-lepton
symmetry. Note that although the group (\ref{eq:G4V}) has the vector form 
the interaction (\ref{eq:lagrVdl}) in general case is not purely vectorlike 
because of the possible difference of the mixing matrices in (\ref{eq:fmix}) 
for the left and for the right quarks and leptons.  The particular case 
of the pure vector interaction in (\ref{eq:lagrVdl}) with $K^{L}_2=K^{R}_2$ has been 
considered in \cite{KM1,KM2}.     

The Higgs mechanism of the quark lepton mass splitting of MQLS-model gives for Yukawa coupling constants 
 the expressions 
\begin{eqnarray}
&&(h^{L,R}_m)_{pi}= h^{L,R}_{pi} c^{(\mp)}_m , 
\label{eq:hmconst}
\end{eqnarray}
\begin{eqnarray}
h^{L,R}_{pi} = -\sqrt{ 3/2} \frac{1}{\eta \sin\beta}
\Big [  m_{d_p} (K_2^{L,R})_{pi} -
(K^{R,L}_2)_{pi}m_{l_i} \Big ]
\label{eq:hconst}
\end{eqnarray}
where $\eta$ is the SM VEV, $\beta$ is the two Higgs doublet mixing angle of the model, 
$m_{d_p}, m_{l_i}$ are the quark and lepton masses 
and $c^{(\mp)}_m$ are the scalar leptoquark mixing parameters in~(\ref{eq:mixS}).

We have calculated the contributions of the scalar and vector leptoquarks into 
the decays $K_L^0 \to e^{\mp} \mu^{\pm}$ and $ B^0 \to e^{\mp} \tau^{\pm} $ 
with accounting also the gluonic corrections of one loop approximation. 
The amplitudes of the decays under considerations in tree approximation 
are calculated in a standard manner and after Firz transformations these amplitudes 
 depend on the matrix elements of the pseudoscalar and axial quark currents 
which are parametrized by the form factors $f_{K^0}$, $f_{B^0}$ of $K^0$, $B^0$ mesons as 
\begin{eqnarray}
&&\langle 0| \bar{s} \gamma^{\mu} \gamma^{5} d| K^0(p) \rangle = i f_{K^0} p^{\mu}, \;\;\;\;\; 
\langle 0| \bar{s} \gamma^{5} d| K^0(p) \rangle = - \, i \, \bbar{m}_{K^0}  f_{K^0},  
\label{eq:fpK}
\\
&&\langle 0| \bar{b} \gamma^{\mu} \gamma^{5} d| B^0(p) \rangle = i f_{B^0} p^{\mu}, \;\;\;\;\; 
\langle 0| \bar{b} \gamma^{5} d| B^0(p) \rangle = - \, i \, \bbar{m}_{B^0}  f_{B^0},  
\label{eq:fpB}
\end{eqnarray}
where $p_{\mu}$ is 4-momentum of the decaying meson and 
\begin{eqnarray}
&&\bbar{m}_{K^0}= m_{K^0}^2/(m_{s}+m_{d}) , \,\,\,\,\,\;\; \bbar{m}_{B^0}= m_{B^0}^2/(m_{b}+m_{d}) .  
\label{eq:mKBT}
\end{eqnarray}

The calculations and analysys of the one loop gluonic corrections to the amplitudes of the 
$K_L^0 \to e^{\mp} \mu^{\pm}$ and $ B^0 \to e^{\mp} \tau^{\pm} $ decays in the case of 
the scalar leptoquark exchange showed that in the leading logarithm approximation 
these corrections give rise to the enhancement of the matrix elements of the pseudoscalar 
quark currents 
by the factors 
\begin{eqnarray}
&&R_{K^0}^S =  R_{K^0}(\mu_{K^0},\mu_0^S), \,\,\,\,\,    R_{B^0}^S =  R_{B^0}(\mu_{B^0},\mu_0^S)
\label{eq:RSKB}
\end{eqnarray} 
depending on mass scale $ \mu_0^S $ at which the Yukawa coupling constants~(\ref{eq:hconst}) 
are defined and on mass scales $ \mu_{K^0} $, $ \mu_{B^0} $ at which the decays occur. 
This dependence can be described as 
\begin{eqnarray}
&& R_{K^0}(\mu_{K^0},\mu_0) = R(\mu_{K^0},m_c; 3) R(m_c,m_b; 4) R(m_b,m_t; 5) R(m_t,\mu_0; 6), 
\label{eq:RK0}
\\
&& R_{B^0}(\mu_{B^0},\mu_0) = R(\mu_{B^0},m_t; 5) R(m_t,\mu_0; 6) 
\label{eq:RB0}
\end{eqnarray} 
with 
\begin{eqnarray}
&& R(\mu_1,\mu_2; n_f) = \big \lbrack \alpha_{st}(\mu_1)/\alpha_{st}(\mu_2) \big \rbrack ^{4/b(n_f)},  
\label{eq:Rmu1mu2}
\end{eqnarray} 
where $ \alpha_{st}(\mu) $ is the strong coupling constant at mass scale $\mu$,  
$ b(n_f) = 11 - (2/3) n_f$, $n_f$ is a number of the active quark flavors.

We will consider below the sums of the widths of the charge conjugated decay modes 
with denoting these sums as
\begin{eqnarray}                                                          
\Gamma(K^0_L \to e \mu ) &\equiv& \Gamma(K^0_L \to e^- \mu^+ )+\Gamma(K^0_L \to e^+ \mu^- )= 
2 \Gamma(K^0_L \to e^- \mu^+ ) ,
\nonumber
\\
\Gamma(B^0 \to e \tau ) &\equiv& \Gamma(B^0 \to e^- \tau^+ )+\Gamma(\ttildeup{B^0} \to e^+ \tau^- )= 
2 \Gamma(B^0 \to e^- \tau^+ ).  \hspace{6mm} 
\nonumber
\end{eqnarray}

With omitting the details of calcilations the final expressions for the widths of the 
$K_L^0 \to e \mu , \; B^0 \to e \tau $ decays induced by the  scalar leptoquarks 
$ S_1, S_2 $ with mixing~(\ref{eq:mixS2}) and with account the one loop gluonic corrections 
for the case of zero fermion mixing 
\begin{eqnarray}
K_2^{L}~=~K_2^{R}~=~I 
\label{eq:fmix0}  
\end{eqnarray}
can be presented in the form    
\begin{eqnarray}
&& 
\Gamma_S(K^0_L \to e \mu ) = 
\frac{m_{K^0} f_{K^0}^2 h_1^2 h_2^2  }{256\pi} \bigg (1-\frac{m_{\mu}^2}{m_{K^0}^2} \bigg )^2 \times
\nonumber\\
&& 
\times \Bigg \{ \bigg \lbrack  m_{\mu} \langle \frac{1}{m^2_S} \rangle^{L} - 
R_{K^0}^S \bbar{m}_{K^0} c s (\frac{1}{m_{S_1}^2} - \frac{1}{m_{S_2}^2}) \bigg \rbrack^2 + 
L \leftrightarrow R  \Bigg \} ,   
\label{eq:gemuK0S}  
\end{eqnarray}
\begin{eqnarray}
&& 
\Gamma_S(B^0 \to e \tau ) = 
\frac{m_{B^0} f_{B^0}^2 h_1^2 h_3^2  }{128\pi} \bigg (1-\frac{m_{\tau}^2}{m_{B^0}^2} \bigg )^2 \times
\nonumber\\
&& 
\times \Bigg \{ \bigg \lbrack  m_{\tau} \langle \frac{1}{m^2_S} \rangle^{L} - 
R_{B^0}^S \bbar{m}_{B^0} c s (\frac{1}{m_{S_1}^2} - \frac{1}{m_{S_2}^2}) \bigg \rbrack^2 + 
L \leftrightarrow R  \Bigg \}    
\label{eq:getauB0S}  
\end{eqnarray}
where 
\begin{eqnarray}
&&h_p = -\sqrt{ 3/2} \frac{1}{\eta \sin\beta} (  m_{d_p} - m_{l_p} ) ,  
\label{eq:hconst2}
\\
&&\langle \frac{1}{m^2_S} \rangle^{L}= \frac{s^2}{m_{S_1}^2}+\frac{c^2}{m_{S_2}^2} \, , \,\,\,\, 
\langle \frac{1}{m^2_S} \rangle^{R}= \frac{c^2}{m_{S_1}^2}+\frac{s^2}{m_{S_2}^2} 
\label{eq:1mS2LR}
\end{eqnarray}
and the relations 
$K^0=(\tilde{s}d)$, $\ttildeup{K^0}=(\tilde{d}s)$, $K^0_L=( (\tilde{s}d)+(\tilde{d}s) )/\sqrt{2}$, $ B^0=(\tilde{b}d)$ 
have been taken into account.

The widths of the decays
\begin{eqnarray}
K^0_L \to l^+_i l^-_j  
\label{eq:decayijK0}  
\end{eqnarray}
with $i,j=1,2, \, l^{\pm}_i=e^{\pm}, \mu^{\pm}$ 
induced by the vector leptoquarks 
with neglect of electron and muon masses $ (m_e, m_{\mu} \ll R_{K^0}^V \bbar{m}_{K^0}) $  
in the case of the general fermion mixing
can be written as  
\begin{eqnarray}
\Gamma_V(K^0_L \to l^+_i l^-_j) = 
 \frac{m_{K^0} \pi \alpha_{st}^2 f_{K^0}^2 \bbar{m}_{K^0}^2 (R_{K^0}^V)^2 }{4m_V^4} 
\,\, \varkappa_{ij}^2 .     
\label{eq:gijK0V}  
\end{eqnarray} 
Here the factor $R_{K^0}^V = R_{K^0}(\mu_{K^0},M_c)$ accounts the gluonic corrections 
to the pseudoscalar quark current and is defined by equation~(\ref{eq:RK0}) with 
mass scale $M_c$ of the four color symmetry breaking  
and the factors 
\begin{eqnarray}
\varkappa_{ij} &=& \sqrt{ (|\varkappa^{L}_{ij}|^2 + |\varkappa^{R}_{ij}|^2)/2 }    
\label{eq:kappaijK0}  
\end{eqnarray} 
with
\begin{eqnarray}
\varkappa^{L,R}_{ij} &=& (K^{L,R}_2)_{2i}\,\,\starupp{(K^{R,L}_2)_{1j}} + 
(K^{L,R}_2)_{1i}\,\,\starupp{(K^{R,L}_2)_{2j}}  
\label{eq:kappaLRijK0}  
\end{eqnarray} 
account the fermion mixing of the general form. 

In particular, for the total width of the $K_L^0 \to e^{\mp} \mu^{\pm}$ decays 
we have 
\begin{eqnarray}
\Gamma_V(K^0_L \to e \mu ) = 
 \frac{m_{K^0} \pi \alpha_{st}^2 f_{K^0}^2 \bbar{m}_{K^0}^2 (R_{K^0}^V)^2 }{2m_V^4} 
\,\, \varkappa_{21}^2      
\label{eq:gemuK0V}  
\end{eqnarray}
where the mixing factor $\varkappa_{21}$ depends on the fermion mixing 
via~(\ref{eq:kappaijK0}), (\ref{eq:kappaLRijK0}) and can be varyed in the region 
$0 \le  \varkappa_{21} \le 1$.
In particular case of $K^{L}_2=K^{R}_2$ the factor $\varkappa_{21}$ reproduces 
the corresponding factor of refs.~\cite{KM1,KM2}.  
In the case~(\ref{eq:fmix0}) of zero fermion mixing $\varkappa_{21}=1$ 
and the decays $K_L^0 \to e^{\mp} \mu^{\pm}$ 
are the only possible decays of type (\ref{eq:decayijK0}) with the total width (\ref{eq:gemuK0V}) 
at $\varkappa_{21}=1$. In this case the width (\ref{eq:gemuK0V}) coincides with that of ref.~\cite{VW}.

The total width of the $B^0 \to e^{\mp} \tau^{\pm}$ decays induced by the vector leptoquarks 
with neglect of electron mass $ (m_e \ll R_{B^0}^V \bbar{m}_{B^0}, \, m_{\tau} ) $  
in the case of zero fermion mixing~(\ref{eq:fmix0}) can be written as  
\begin{eqnarray}
&&\Gamma_V(B^0 \to e \tau ) = 
\frac{m_{B^0}\pi \alpha_{st}^2  f_{B^0}^2}{m_V^4}  
( R_{B^0}^V \bbar{m}_{B^0} - m_{\tau}/2)^2 \bigg (1-\frac{m_{\tau}^2}{m_{B^0}^2} \bigg )^2
\label{eq:getauB0V}  
\end{eqnarray} 
where the factor $R_{B^0}^V =  R_{B^0}(\mu_{B^0},M_c)$ accounts the gluonic corrections.

For comparision with the case of the vector leptoquarks $V$ we have also calculated 
the widths of $K_L^0 \to e^{\mp} \mu^{\pm}$ and $ B^0 \to e^{\mp} \tau^{\pm} $ decays 
induced by the left $(V^L)$ and right $(V^R)$ chiral gauge leptoquarks. 

The interaction of the chiral gauge leptoquarks with fermions can be written as 
\begin{eqnarray}
  \emph{L}_{Vdl} &=& \frac{g^L_4}{\sqrt{2}} (\bar{d}_{p \alpha} 
[ (K^L_2)_{pi}\gamma^{\mu}P_L] l_i)V^L_{\alpha \mu} +  
\frac{g^R_4}{\sqrt{2}} (\bar{d}_{p \alpha} 
[ (K^R_2)_{pi}\gamma^{\mu}P_R] l_i)V^R_{\alpha \mu} + h.c. , \hspace{5mm} 
\label{eq:lagrVLRdl}
\end{eqnarray}
where  $g^{L}_4$, $g^{R}_4$ are the gauge coupling constants of the group~(\ref{eq:G4LR}) or~(\ref{eq:G4SLR}) 
which are related to strong coupling constant by the equation 
\begin{eqnarray}
g^L_4 g^R_4 / \sqrt{(g^L_4)^2+(g^R_4)^2} = g_{st} . 
\label{eq:gLgRgst}
\end{eqnarray}

The resulted total widths of 
the $K_L^0 \to e \mu , \; B^0 \to e \tau $ decays with account of the contributions 
of the chiral gauge leptoquarks $ V^L, V^R $ with neglecting the electron mass can be written as   
\begin{eqnarray}
\Gamma_{V^{LR}}(K^0_L \to e \mu ) =  \frac{m_{K^0}f_{K^0}^2m_{\mu}^2} {64\pi} 
\bigg (1-\frac{m_{\mu}^2}{m_{K^0}^2} \bigg )^2 
\bigg \lbrack \frac{(g^L_4)^4}{4m_{V^{L}}^4} |\varkappa'^L_{21}|^2 + 
L \leftrightarrow R \bigg \rbrack , \hspace {8mm}   
\label{eq:gemuK0VLR}  
\end{eqnarray}
\begin{eqnarray}
\Gamma_{V^{LR}}(B^0 \to e \tau ) &=&  \frac{m_{B^0}f_{B^0}^2m_{\tau}^2} {32\pi} 
\bigg (1-\frac{m_{\tau}^2}{m_{B^0}^2} \bigg )^2 
\bigg \lbrack \frac{(g^L_4)^4 }{4m_{V^{L}}^4} |k'^L_{31}|^2 + L \leftrightarrow R  \bigg \rbrack 
\label{eq:getauB0VLR}  
\end{eqnarray}
where the parameters 
\begin{eqnarray}
\varkappa'^{L,R}_{ij} &=& (K^{L,R}_2)_{2i}\,\,\starupp{(K^{L,R}_2)_{1j}} + 
(K^{L,R}_2)_{1i}\,\,\starupp{(K^{L,R}_2)_{2j}} ,    
\label{eq:kappa1LRijK0}  
\end{eqnarray} 
\begin{eqnarray}
k'^{L,R}_{ij} &=& (K^{L,R}_2)_{3i}\,\,\starupp{(K^{L,R}_2)_{1j}}  
\label{eq:k1LRijB0}  
\end{eqnarray}
account the effects of the general fermion mixing. 
The widths~(\ref{eq:gemuK0VLR}),~(\ref{eq:getauB0VLR}) depend on the matrix elements 
of the axial currents in~(\ref{eq:fpK}),~(\ref{eq:fpB}) and as it has been shown 
by the corresponding analysys the gluonic corrections of the leading logarithm approximation 
in this case are absent.   

We have numericaly analysed 
the widths~(\ref{eq:gemuK0S}),~(\ref{eq:getauB0S}),~(\ref{eq:gemuK0V}),~(\ref{eq:getauB0V}) and 
(\ref{eq:gemuK0VLR}),~(\ref{eq:getauB0VLR}) 
in dependence on the leptoquark masses. 
The Yukawa coupling constants~(\ref{eq:hconst}) are defined by the quark masses at 
the mass scale $\mu_0^S = \eta$ of order  of the mass scale of the SM symmetry breaking,  
$\eta = 250 \, GeV $ is the SM VEV.     
The gauge coupling constants are related 
to $g_{st}(M_c)$ at $\mu_0^V=M_c=1000 \, TeV$, in the case of chiral gauge leptoquarks 
we assume also that $g^L_4=g^R_4(=\sqrt{2}g_{st})$. 
The mass scales of the decaying particles are chosen at $ \mu_{K^0}=1 \, GeV $ and $ \mu_{B^0}=m_{B^0} $ 
for $K_L^0$ and $B^0$ decays respectively and these mass scales are also used for defining the quark masses 
in~(\ref{eq:fpK}),~(\ref{eq:fpB}),~(\ref{eq:mKBT}).   
With implying the isotopic symmetry we use the value of the form factor 
$ f_{K^0}= f_{K^+}= 160 \, MeV$~\cite{PDG06} and the central value of   
$f_{B^0}= f_{B^-}=229^{+36}_{-31}(stat)^{+34}_{-37}(syst) \, MeV$~\cite{Ikado}.  
We use also the known life times of $K_L^0$ and $B^0$ mesons~\cite{PDG06}.         

Fig.1 shows the branching ratio of $K^0_L \to e \mu$ decay 
for the case of zero fermion mixing~(\ref{eq:fmix0})  
in dependence on $m_S \sin \beta$ where $m_S = m_{S^L}$, $m_{S^R}$, $m_{S^S}$, $m_{S^P}$ 
are the scalar leptoquark masses for two cases of the scalar leptoquark mixings  
1) $ sin\theta=0, \, S^L=\starup{S_2^{(-)}}, \, S^R=S_2^{(+)} $  
and 2) $ sin\theta=1/\sqrt{2}, \, S^S=(S_2^{(+)}+\starup{S_2^{(-)}})/\sqrt{2} , \, 
S^P=(S_{2}^{(+)}-\starup{S_2^{(-)}})/\sqrt{2} $. 
The horizontal dashed line shows the experimental upper limit~\cite{PDG06} 
\begin{eqnarray}
Br(K^0_L \to e \mu)<4.7\cdot 10^{-12} .  
\label{eq:BrKexp}  
\end{eqnarray}

The first two curves correspond to the case of the chiral scalar leptoquark mass states for 
$m_S=m_{S^L} \ll m_{S^R}$ (curve $1)$) and for $m_S=m_{S^L}=m_{S^R} $ (curve $2)$).  
The next two curves correspond to the case of the leptoquark mass states of the scalar type 
$m_S=m_{S^S} \ll m_{S^P} $ (curve $3)$) and of the pseudoscalar one 
$m_S=m_{S^P} \ll m_{S^S} $(curve $4)$).
As seen in all the cases the lower mass limits for the scalar leptoquarks 
are small, of order or below the mass limits from the direct search for scalar leptoquarks. 

More exactly, from~(\ref{eq:gemuK0S})  and~(\ref{eq:BrKexp}) we obtain 
the next scalar leptoquark mass limits 
\begin{eqnarray}
 m_{S^L}, m_{S^R} > 15/\sin \beta \,\,\,\, GeV ,  
\label{eq:mlimSLR}
\\  
 m_{S^S}, m_{S^P} > 60/\sin \beta \,\,\,\, GeV. 
\label{eq:mlimSSP}  
\end{eqnarray}
For a validity of a pertubation theory the Yukawa coupling constants 
including those for $t$-quark should be sufficiently small, 
which implys that $0.2 \le \sin{\beta} \le 1$.  
So, the lower mass limits~(\ref{eq:mlimSLR}) for chiral scalar leptoquarks are 
below the mass limits from the direct search for scalar leptoquarks and those~(\ref{eq:mlimSSP})
 for the scalar leptoquarks of scalar or pseudoscalar types can be of order or below 
their direct mass limits.  

Fig.2 shows the branching ratios of $K^0_L \to e \mu$ decay 
as a function of the ratios $ m_{V}/\sqrt{\varkappa_{21}}, \, m_{V^L}/\sqrt{|\varkappa'^L_{21}|} $ 
of the  masses $m_{V}, \, m_{V^L}$ of 
the vector and chiral gauge leptoquarks to the corresponding fermion mixing parameters.  
The branching ratio of $K^0_L \to e \mu$ decay      
as a function of the ratio $ m_{V}/\sqrt{\varkappa_{21}}$ 
is shown for the cases with account of gluonic corrections (curve $1$) 
and (for comparision) with neglecting them (dashed line).  
The curve 2 shows the branching ratio of $K^0_L \to e \mu$ decay 
in dependence on the ratio $ m_{V^L}/\sqrt{|\varkappa'^L_{21}|} $ 
with assuming for definiteness that $m_{V^L} \ll m_{V^R}$.  
As seen, in the case of zero fermion mixing ($ \varkappa_{21}=|\varkappa'^L_{21}|=1 $) 
 the mass limits for the gauge leptoquarks 
are essentially  more stringent than those for 
the scalar leptoquarks. 

Using~(\ref{eq:gemuK0V}),~(\ref{eq:gemuK0VLR}) and~(\ref{eq:BrKexp}) we obtain 
from $K_L^0 \to e^{\mp} \mu^{\pm}$ decays the next gauge leptoquark mass limits 
\begin{eqnarray}
&& m_V >  \sqrt{\varkappa_{21}} \,\,\, 2000 \,\,\,\, TeV ,
\label{eq:mlimVK0}
\\  
&& m_{V^L} > \sqrt{|\varkappa'^L_{21}|} \,\,\, 260 \,\,\,\, TeV 
\label{eq:mlimVLK0}  
\end{eqnarray}
for the vector and chiral (with assuming $m_{V^L} \ll m_{V^R}$ ) gauge leptoquarks.   
The mass limit~(\ref{eq:mlimVK0}) for the vector leptoquarks coprrespons 
to the current date~(\ref{eq:BrKexp}) and it is more stringent than the known one~\cite{KM1,KM2,PDG06} 
whereas~(\ref{eq:mlimVLK0}) is the new mass limit for the chiral gauge leptoquarks.  

It interesting to note that the simultaneous account of the gauge and scalar leptoquark contributions 
into decays under consideration can weaken the leptoquark mass limits due the possible destructive 
interference of these contributions. For example such destructive interference takes place between 
the contributions of the vector leptoquarks $V$ and of the scalar leptoquarks $S^S$ of the scalar type.  
For $m_{S^S}$ from~(\ref{eq:mlimSSP}) this interference reduces the mass limit for vector leptoquarks 
from~(\ref{eq:mlimVK0}) to $m_V > \sqrt{\varkappa_{21}} \,\,\, 1400 \,\,\,\, TeV$. 
      
The current experimental limit on the total branching ratio of 
$B^0 \to e^{\mp} \tau^{\pm}$ decays~\cite{PDG06} 
\begin{eqnarray}
Br(B^0 \to e \tau)<1.1\cdot 10^{-4}
\label{eq:BrBexp}  
\end{eqnarray} 
gives the relatively weak limits on the leptoquark masses. 

The lower mass limits from $B^0 \to e^{\mp} \tau^{\pm}$ decays 
for the scalar leptoquarks resulted from~(\ref{eq:getauB0S}),~(\ref{eq:BrBexp}) are only 
of order of a few GeV, i.e. they are essentially weaker than 
those~(\ref{eq:mlimSLR}),~(\ref{eq:mlimSSP}) from $K_L^0 \to e^{\mp} \mu^{\pm}$ decays. 

Using~(\ref{eq:getauB0V}),~(\ref{eq:getauB0VLR}) and~(\ref{eq:BrBexp}) we have obtained 
from $B^0 \to e^{\mp} \tau^{\pm}$ decays 
the next mass limits for the gauge leptoquarks   
\begin{eqnarray}
&& m_V > 9.3 \,\,\,\, TeV ,
\label{eq:mlimVB0}
\\  
&& m_{V^L} > \, \sqrt{|k'^L_{31}|} \,\,\,  2.8 \,\,\,\, TeV.  
\label{eq:mlimVLB0}  
\end{eqnarray} 
The mass limit~(\ref{eq:mlimVB0}) correspond to the case~(\ref{eq:fmix0}) 
and can be lowered by the fermion mixing.  
The mass limits~(\ref{eq:mlimVB0}), (\ref{eq:mlimVLB0}) are weaker than 
those~(\ref{eq:mlimVK0}),~(\ref{eq:mlimVLK0}) from $K_L^0 \to e^{\mp} \mu^{\pm}$ decays  
nevertheless they are of interest as the new independent ones. 

It is worth to note that unlike the $K_L^0$-decays~(\ref{eq:decayijK0}) 
in the case of $B^0$-meson all the decays 
\begin{eqnarray}
B^0 \to l^+_i l^-_j  
\label{eq:decayijB0}  
\end{eqnarray}
with $i,j=1,2,3, \, l^{\pm}_i=e^{\pm}, \mu^{\pm}, \tau^{\pm}$ 
are allowed. The search for the decays~(\ref{eq:decayijB0}) and the measurements of their 
branching ratios will give the possibility to set the leptoquark mass limits with account 
of the fermion mixing of the general form and are of interest.   

In conclusion we resume the results of the work. 
The contributions of the scalar and gauge leptoquarks into widths of
the $K^0_L \to e^{\mp} \mu^{\pm}$, $B^0 \to e^{\mp} \tau^{\pm}$
decays are calculated in the models with the vectorlike and chiral four color symmetry 
and with the Higgs mechanism of the quark-lepton mass splitting. 
From the current data on leptonic $K^0_L$ decays 
the mass limits~(\ref{eq:mlimSLR}),~(\ref{eq:mlimSSP}),~(\ref{eq:mlimVLK0})  
for scalar and chiral leptoquarks as well as the updated  
mass limit~(\ref{eq:mlimVK0}) for vector leptoquarks are obtained. 
The gauge leptoquark mass limits~(\ref{eq:mlimVB0}),~(\ref{eq:mlimVLB0}) from 
the current data on $B^0 \to e^{\mp} \tau^{\pm}$ decays are also obtained, 
which occur to be essentially weaker than those from $K^0_L \to e^{\mp} \mu^{\pm}$ decays.     
It is shown that in all the cases considered the scalar leptoquark mass limits 
(unlike the gauge leptoquark ones) are weak, 
of order or below their direct mass limits.   
The search for such scalar leptoquarks at LHC and the further search for    
the leptonic decays $ B^0 \to l^+_i l^-_j $ are of interest.

\vspace{3mm} {\bf Acknowledgments}

The work was partially supported by the Russian Foundation for
Basic Research under grant 04-02-16517-a.

\newpage

{\Large\bf Figure captions}

\bigskip

\begin{quotation}

\noindent
Fig. 1.  Branching ratio of $K^0_L \to e \mu$ decay 
in dependence on $m_{S} \sin \beta $ for \hspace{20mm} 
$1)m_S=m_{S^L} \ll m_{S^R}, $ \,  
$2)m_S=m_{S^L}=m_{S^R},     $ \,
$3)m_S=m_{S^S} \ll m_{S^P}, $ \,
$4)m_S=m_{S^P} \ll m_{S^S}. $

\noindent
Fig. 2. Branching ratio of $K^0_L \to e \mu$ decay 
in dependence on the ratio $m_{V}/\sqrt{\varkappa_{21}}$ of the vector leptoquark mass  
to the fermion mixing parameter $\sqrt{\varkappa_{21}}$ 
with account of gluonic corrections (curve $1$) and with neglecting them (dashed line) 
and on the ratio $m_{V^L}/\sqrt{|\varkappa'^L_{21}|}$ of the left chiral leptoquark mass  
to the fermion mixing parameter $\sqrt{|\varkappa'^L_{21}|}$ (curve $2$).   


\end{quotation}

\newpage

\begin{figure}[htb]
\vspace*{0.5cm}
\centerline{\epsfxsize=0.8\textwidth\epsffile{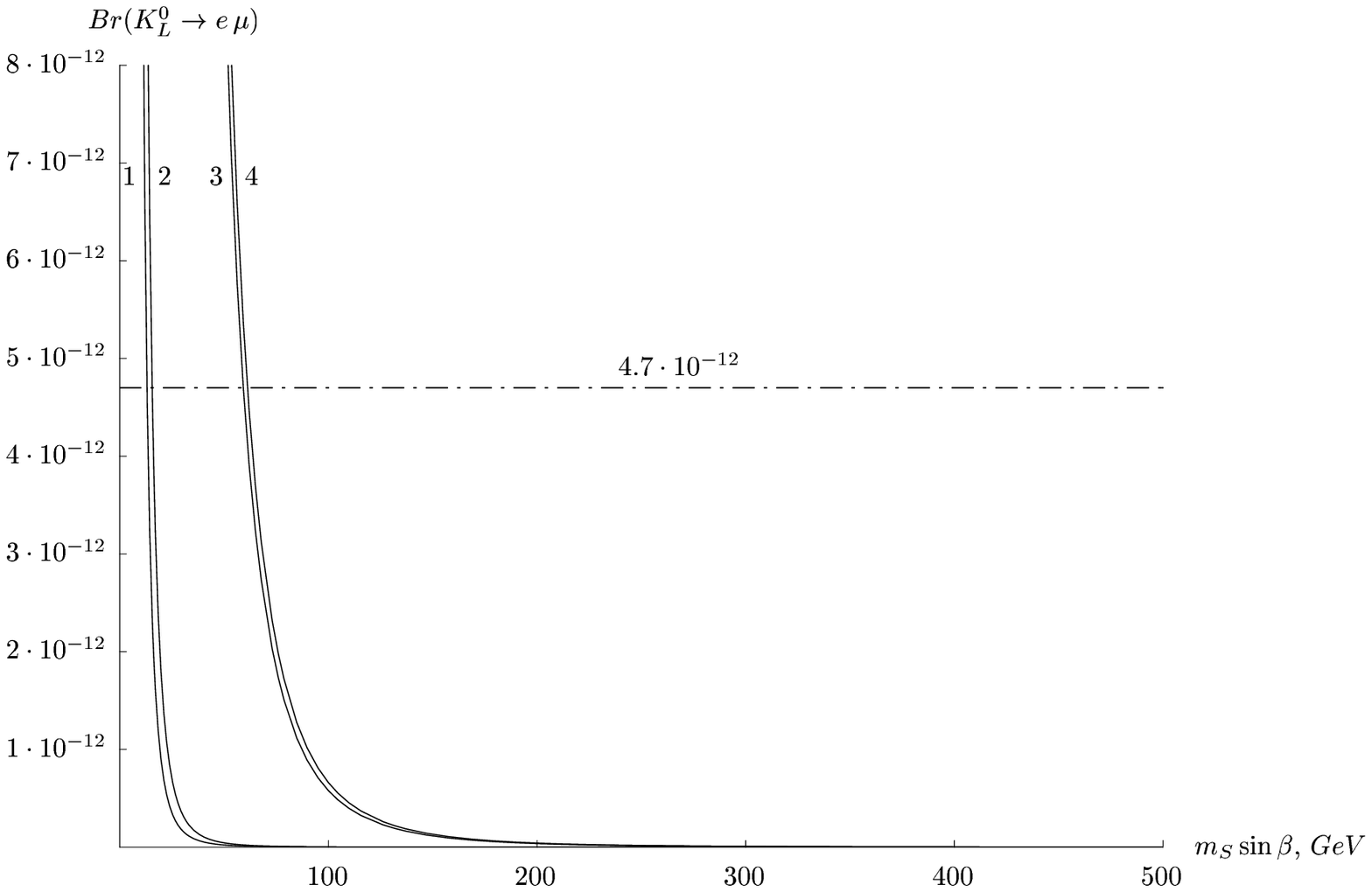}} 
\label{fig:BrKS} 
\end{figure}

\vfill \centerline{A.D.~Smirnov, Modern Physics Letters A}

\centerline{Fig. 1}

\newpage

\begin{figure}[htb]
\vspace*{0.5cm}
 \centerline{\epsfxsize=0.8\textwidth \epsffile{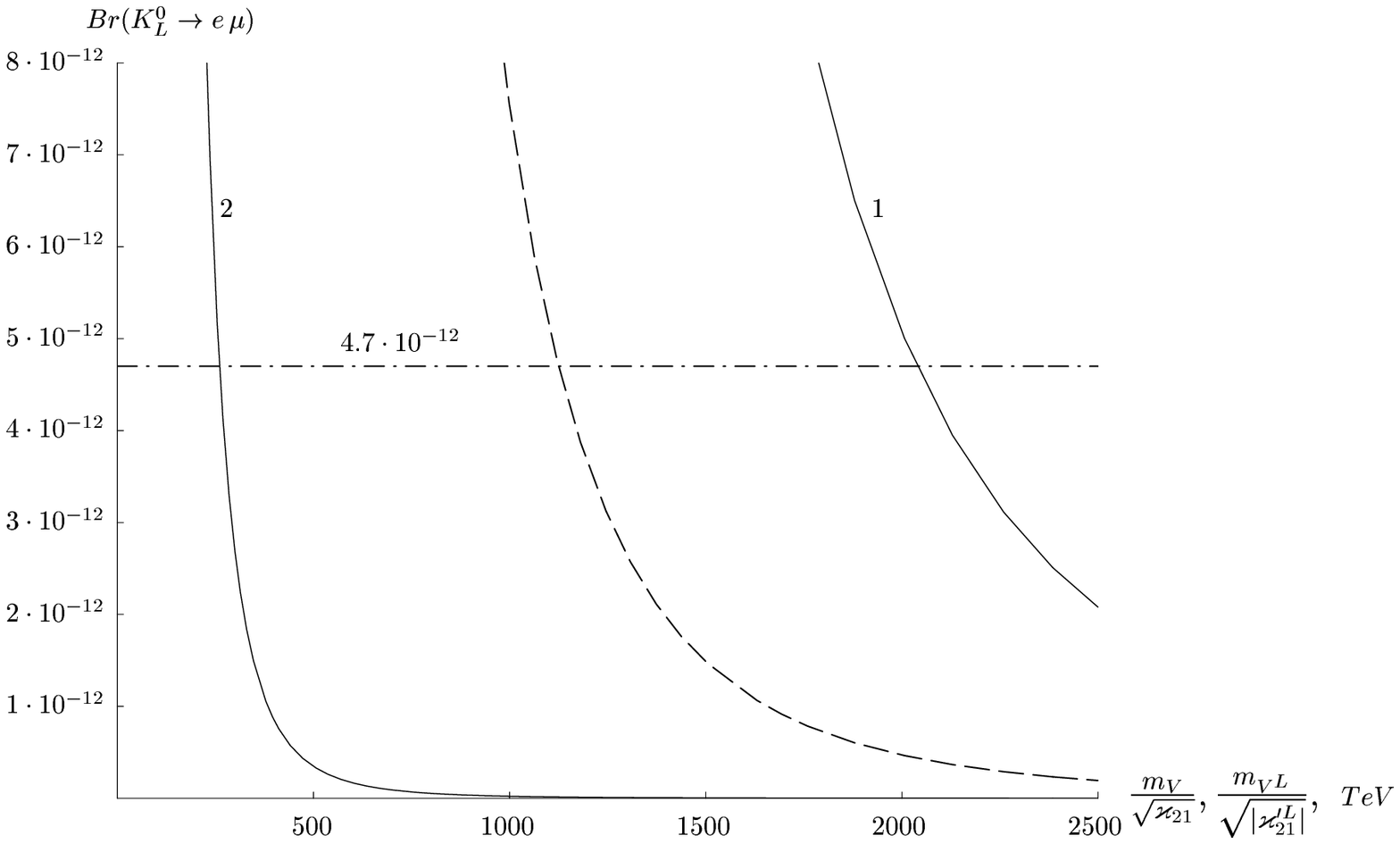}} 
\label{fig:BrKV} 
\end{figure}

\vfill \centerline{A.D.~Smirnov, Modern Physics Letters A}

\centerline{Fig. 2}


\vspace{3mm}


\newpage
\vspace{-5mm}
\begin{thebibliography}{99}
\vspace{-2mm}



\bibitem{PS}
    J.C.~Pati, A.~Salam, Phys.~Rev. D10~(1974)~275.
\bibitem{AD1}
    A.D.~Smirnov, Phys.~Lett. B346~(1995)~297.
\bibitem{AD2}
        A.D.~Smirnov, 
        Phys. At. Nucl. 58~(1995)~2137.
\bibitem{VW}
    G.~Valencia, S.~Willenbrock, Phys.~Rev. D50~(1994)~6843.
\bibitem{KM1}
    A.~V.~Kuznetsov, N.~V.~Mikheev, Phys.~Lett. B329~(1994)~295.
\bibitem{KM2}
    A.~V.~Kuznetsov, N.~V.~Mikheev, 
        Phys. At. Nucl. 58~(1995)~2228.
\bibitem{BRW}
        W.~Buchm\"uller, R.~R\"uckl, D.~Wyler, Phys.~Lett.
        B191~(1987)~442.
\bibitem{PDG06}
        Particle Data Group, W.-M.~Yao et~al., J.~Phys. G33~(2006)~1.
\bibitem{ADPv}
        A.~V.~Povarov, A.~D.~Smirnov,  
        Phys. At. Nucl. 64~(2001)~74.
\bibitem{AD3}
        A.~D.~Smirnov, Phys.~Lett.~B513 (2002) 237.
\bibitem{PovSm2} 
        A.V. Povarov and A.D. Smirnov,            
        Phys. At. Nucl. 66~(2003)~2208.
\bibitem{Ikado} 
        K.~Ikado et~al., Phys.~Rev.~Lett. 97~(2006)~251802;  hep-exp/ 0604018. 



\end{thebibliography}
\end{document}